# "Phase Freezeout" in Isentropically Expanding Matter

Iosilevskiy I.L.

*Joint Institute for High Temperature (Russian Academy of Science) Moscow, Russia*
ilios@orc.ru

**Introduction**

Features of isentropic (iso-$S$) expansion of warm dense matter (WDM), which was created by intense energy flux (strong shock compression or instant isochoric heating by laser or/and heavy ion beam) are under discussion in situation when:

($i$) – thermodynamic trajectory of such expansion crosses binodal of liquid-gas phase transition (i.e. boiling curve), and

($ii$) – expansion within the two-phase region is going on along *equilibrium* branch (not metastable one) of the two-phase mixture isentrope. It is presumed that such equilibrium two-phase mixture ("mixed phase") is *highly dispersed* so that two standard approaches are applicable: LTE – local thermodynamic equilibrium approximation and LHA – local hydrodynamic approximation.

**Liquid layer at isentropic expansion of semi-infinite WDM in plane geometry**

It is known for the plane geometry [1] (Anisimov, Inogamov and Rethfeld) that there appears extended *zone of uniformity* (layer) with *constant thermodynamic parameters* in isentropically expanding material, which corresponds just to the matter in *boiling liquid state* on binodal. This comes true because of sharp break of expansion-isentrope in *P-V* plane just at the boiling point (i.e. at binodal) because off high jump of sound velocity at this point. There are three parts of isentrope, that are realized in this scenario of expansion (see fig. 1,2): - part of isentrope in condensed phase (*A*), part of isentrope in two-phase region (*C*) and zone corresponding to the boiling liquid (*B*). It should be stressed that this zone of the boiling liquid corresponds only to one point on the phase diagram of substance (binodal B at fig.1), and at the same time to extensive layer of final thickness on the dynamic $x - t$ diagram (fig.2).

It is known that hydrodynamics of isentropic expansion for semi-infinite WDM sample is self-similar (its description does not contain scale factor). It means that spatial profiles of all parameters (density, pressure, temperature etc) depends on $x$ and $t$ only in combination $x/t$. It should be emphasized that because of this self-similarity this zone of boiling liquid has remarkable property – it contains *finite* and *fixed* portion of whole mass of isentropically expanding material. For example, according to numerical calculation of [2] this portion can be estimated as of order of unity.

This remarkable property makes it possible (at least formally) to discuss this type of isentropic WDM expansion as a tool for *generation* (and subsequent diagnostics) of *extended uniform* state of the matter *exactly* on its *binodal* (and even in *critical point*!) for the case when parameters of this binodal (and/or thermophysical properties on it) are poorly known (for example [3]). Just for this reason it is natural to use for this regime of expansion the term: - "*phase freezeout*". It is similar to the terms "chemical freezeout" and

"kinetic freezeout", which are widely used in interpretation of quark-hardon transformations during the expansion of products for ultra-high energy ionic collisions in supercolliders [4].

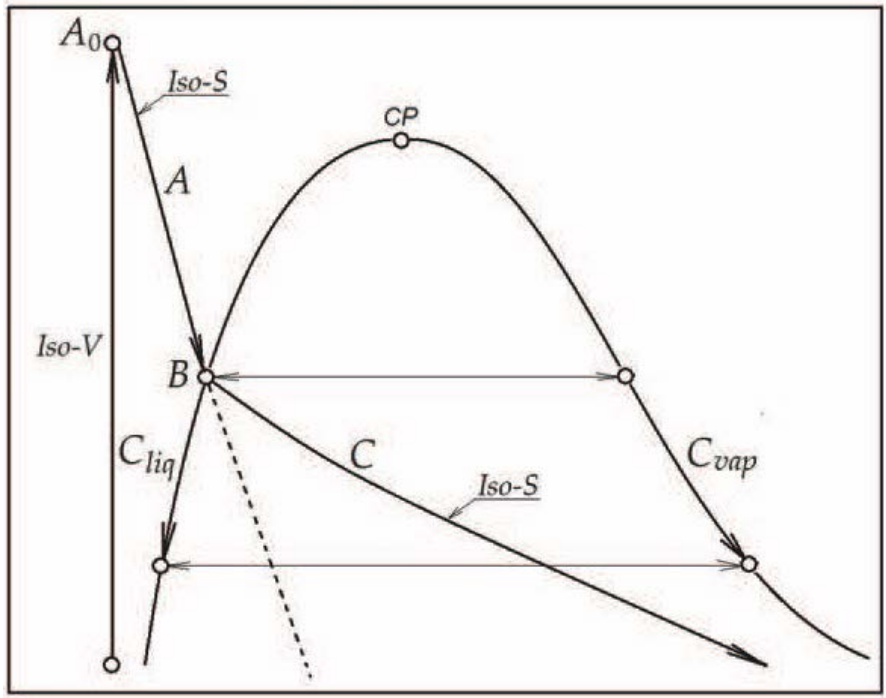

**Figure 1**. Three parts of isentropic expansion at *P-V* phase diagram: *A* – isentrope in condensed phase, *C* – equilibrium branch of isentrope in two-phase region (equilibrium mixture of liquid ($C_{liq}$) and gaseous ($C_{gas}$) phases), and *B* – point of boiling liquid at binodal of liquid-gas phase transition. *CP* – critical point. Dash line – metastable branch of isentrope.

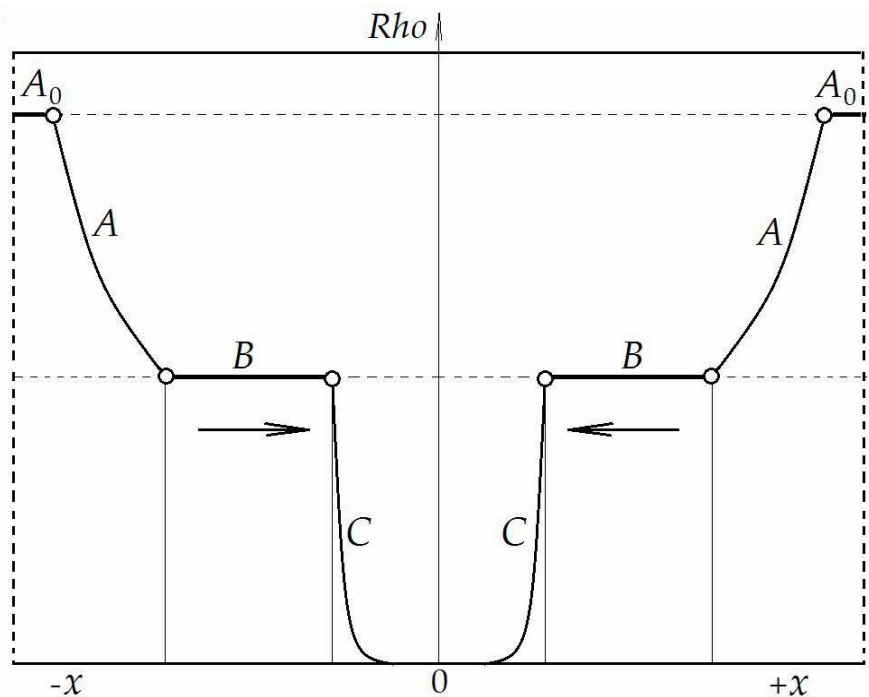

**Figure 2**. Three parts of expansion isentrope for two colliding layers at $\rho$–$x$ diagram: *A* – condensed phase, *C* – two-phase mixture and *B* – zone of boiling liquid at binodal of liquid–gas phase transition.

**Isentropic expansion of finite WDM-layer (slab) in plane geometry.**

Characteristic scale (slab width – $d$) presents in this case. Hydrodynamics of isentropic expansion of WDM-slab into both sides is still self-similar, but only up to the moment of crossover for convergent rarefaction waves in the slab center ($t_0 < t < t_1$). As well as in previous case, three zones of expansion isentrope are realized in expanding substance: – liquid zone (*A*), two-phase one (*C*) and zone of boiling liquid, which corresponds to one *point* at liquid binodals (*B*) at phase diagram (fig.1) and to extensive *layer* on *x-t* diagram (fig.2). After that meeting central density of WDM-slab decreases (during interval $t_1 < t < t_2$) from initial density ($\rho_0$) to the density of boiling liquid ($\rho_b$). Just at this moment expanding substance reaches remarkable combination of two uniform layers of boiling liquid, which are flying away to the opposite sides with velocity, determined by enthalpy difference along the expansion isentrope between initial point $A_0$ and the point *B* at liquid binodal [i.e. $\Delta H = H(\rho_0, S) - H(\rho_b, S)$]. After that moment these two layers continue their movement in opposite direction by the inertia, leaving between themselves the flat cavity filled in with two-phase mixture (foam) with lowered (but positive) pressure, corresponding to descending (two-phase) section of expansion isentrope (part *C* at fig.1). Formally, away of hydrodynamic instability problem, this isentropic flying away of two uniform boiling liquid layers continues unlimitedly with constant velocity and gradual loss of mass of each layer. In maximally simplified approximation, i.e. if one neglect with mass and energy (internal and kinetic ones) of two-phase portions between and outside of the expanding liquid layers, then in accordance with the first thermodynamics law entire difference in the internal energy between the states $A_0\{\rho_0, P_0, T_0\}$ and $B\{\rho_b, P_b, T_b\}$ transforms into kinetic energy of two moving layers of boiling liquid.

**Hypothetical generation of extended uniform zone of boiling liquid via isentropic expansion of stack target.**

Dynamics (and thermodynamics) of described above isentropic decomposition of isolated (single) WDM-slab allows us, at least formally, to discuss hypothetical scenario for generation of extensive and uniform sample of studying condensed material of *arbitrary width* exactly in the state of *boiling liquid*. It could be done with the use of thoroughly positioned *combination* of such slabs when they are instantly, uniformly and volumetrically heated, for example, by intensive heavy ion beam. The idea of using of such colliding layers has been discussed many times (see for example [6]). Prospective of using for special multi-layered condensed target ("stack target") in the regime of the controlled thermal *quasi-isobaric expansion* under energy deposition from heavy ion beam (HIB) with presently available energy input of moderate intensity (~ 1 kJ/g) were discussed qualitatively and quantitatively in [5]. Different more simple variants for realization of such quasi-isobaric regime for thermal expansion of *highly dispersive porous* condensed target, volumetrically heated by HIB from initial porous state up to reaching of the state of continuity were claimed and discussed long before in [6-8]. Possible schemes and perspectives for generation of uniform state of investigated condensed substance in the regime of slow (relatively) heating *along boiling (sublimation) curve* (titled there as "tracing of boiling curve") also were discussed in these papers. Promising alternative way for generation of such uniform and extensive sample of investigated substance just at the boiling (sublimation) condition is presently discussed for the case of perspective future energy deposition high enough for almost instant *quasi-isochoric* heating up to the entropy level as high as entropy in critical point of gas-liquid phase transition in this investigated substance (scheme HIHEX in the HEDgeHOB Collaboration [9]).

The main condition for reaching of such regime is proper choice for two key parameters: first one is ratio of width for slabs $d$ and slots between them $f$ (i.e. porosity ratio – $d/f$), and second – the level of heating the material of stack target, so that two events – impact of colliding liquid layers, moment $t_3$, and disappearance of compressed condensed state between scattering liquid layers (i.e. meeting of the back sides of liquid layers (moment $t_2$) would coincide, i.e. $t_2 = t_3$. In this case just at the moment $t_2 = t_3$ all portions of a substance beyond liquid-gas boundary (binodal): i.e. those in the state of two-phase mixture, and those in the state of still compressed liquid between the scattering boiling liquid layers, would disappear, and entire irradiated target come into the state of the *continuous* and thermodynamically *uniform* combination of scattering an colliding layers of boiling liquid with equal, but differently directed speed. Formally this moment "*X*" is just the moment of desired generation of *uniform* and *extensive* state for investigated substance exactly on the boiling boundary (binodal), in particular, including the most interesting case – generation of such state just at the critical point of gas-liquid phase transition. With given parameters at initial and binodals point, $A\{P_0, \rho_0, T_0\}$ and $B\{P_b, \rho_b, T_b\}$, which strongly depend on the level of HIB energy deposition, proper geometric condition for reaching of discussed regime is determined by proper choice for the geometrical ratio $d/f$ to correspond to the density ratio $\rho_0/\rho_b$.

$$d/f = \rho_0/\rho_b \qquad (1)$$

It should be emphasized that special feature of the discussed regime for isentropic expansion of WDM-material makes it possible to discuss, at least formally, possibility of generation for uniform and extensive sample of investigated substance, which we are interested in, just on the boiling boundary of liquid-gas phase transition (including the region of its critical point) with the size of this zone, being determined by the available size of uniform volumetric heating via high-energy HIB deposition in future accelerator SIS-100 (project FAIR).

**Comments to hydrodynamics of hypothetical "reverberation" regime in adiabatic movement of stack target.**

Hydrodynamics of subsequent movement for all pairs of colliding layers of boiling liquid after the moment of meeting for two counter layers ("*X*" moment) is very complicated. At sufficiently high velocity of this collision one can expect shock wave stopping of both the colliding layers. Thermodynamic state of these colliding liquid layers moves from the point on liquid binodal to corresponding point at Hugoniots of compressed liquid. General hydrodynamic scheme of adiabatic movements for the case of more complicated stack target (as an ensemble of well positioned plates or foils) under instant isochoric heating (for example, by heavy ion beam) could be considered as natural elaboration of mentioned above simple initial idea of phase freeze-out.

In conclusion, one could expect hydrodynamics of such WDM stack to be a descending sequence of alternate isentropic expansions and shock compressions, *etc. etc.* up to the full stoppage at final point at binodal of boiling liquid, including the most interesting case – state in critical point. Indeed, we need more theoretical and numerical efforts for clarification of this complicated hydrodynamics.

**Comments to the problem of quark-gluon fireball adiabatic expansion**

It is interesting to discuss relevance of mentioned above phenomenon – hypothetical stoppage for significant portion of isentropically expanding material just at the boundary of phase transition in connection with the problem of adiabatic (isentropic) expansion of thermalized quark-gluon "fireball", which is created after huge impact of two relativistic ions in ultra-high energy ionic collision in supercolliders (see for example [4] and materials of other CPOD conferences). It seems to be natural to use for this regime of expansion the term: *"phase freezeout"* [11]. Its meaning is similar to the meaning of the terms "chemical freezeout" and "kinetic freezeout", which are widely used in interpretation of quark-hardon transformations during the expansion of products for ultra-high energy ionic collisions in super-colliders [4]. Perspective of such quark-gluon phase freeze-out in form of expanding spherical layer in the state just at the binodal of quark-hadron phase transition could be discussed in connection with isentropic expansion of quark-gluon "fireball", created by mentioned above huge ionic collision.

**Conclusions and perspectives**

It should be emphasized that special feature of the discussed regime for isentropic expansion of WDM-material makes it possible to discuss, at least formally, possibility of generation for uniform and extensive sample of investigated substance, which we are interested in, just on the boiling boundary of liquid-gas phase transition (including the region of its critical point) with the size of this zone, being determined by the available size of uniform volumetric heating via high-energy HIB deposition in future accelerator SIS-100 (project FAIR). It is natural that by one of the most promising methods of diagnostics of this state of investigated material could be considered the method of high energy proton microscopy (HEPM), which is discussed and developed actively in recent years [10].

**Acknowledgements**

This work was supported by the RAS Scientific Program "Physics of extreme states of matter", MIPT Education Center "Physics of High Energy Density Matter" and Extreme Matter Institute – EMMI (Germany)